\documentclass[10pt]{article}

\usepackage{authblk} 
\usepackage[margin=1in]{geometry} 
\usepackage{graphicx} 
\usepackage{url}
\usepackage{amsmath,amssymb} 
\usepackage{authblk}

\usepackage[super,numbers,sort&compress]{natbib}
\setcitestyle{comma,open={},close={}}

\usepackage{lineno} 
\newcommand{\numberthis}{\addtocounter{equation}{1}\tag{\theequation}}

\renewenvironment{abstract}
 {\normalsize
  \begin{center}
    \bfseries \abstractname\vspace{-.5em}\vspace{0pt}
  \end{center}
  \list{}{%
    \setlength{\leftmargin}{.5in}%
    \setlength{\rightmargin}{\leftmargin}%
  }%
  \item\relax}
 {\endlist}

\usepackage{caption}
\begin{document}

\title{\textbf{Quantifying cascading power outages during climate extremes considering renewable energy integration}}

\author[a,d]{Luo Xu\thanks{Corresponding author: luoxu@princeton.edu}}
\author[a,d]{Ning Lin}
\author[b]{H. Vincent Poor}
\author[a]{Dazhi Xi}
\author[c]{A.T.D. Perera}

\affil[a]{\small{Department of Civil and Environmental Engineering, Princeton University, Princeton, NJ, USA}}
\affil[b]{Department of Electrical and Computer Engineering, Princeton University, Princeton, NJ, USA}
\affil[c]{Andlinger Center for Energy and the Environment, Princeton University, Princeton, NJ, USA}
\affil[d]{Center for Policy Research on Energy and the Environment, Princeton University, Princeton, NJ, USA}

\date{}

\maketitle

\begin{abstract}
Climate extremes, such as hurricanes, combined with large-scale integration of environment-sensitive renewables, could exacerbate the risk of widespread power outages. We introduce a coupled climate-energy model for cascading power outages, which comprehensively captures the impacts of evolving climate extremes on renewable generation, and transmission and distribution networks. The model is validated by the 2022 Puerto Rico catastrophic blackout during Hurricane Fiona — the first-ever system-wide blackout event with complete weather-induced outage records. The model presents a novel resilience pattern that was not captured by the present state-of-the-art models and reveals that early failure of certain critical components surprisingly enhances overall system resilience. Sensitivity analysis of various behind-the-meter solar integration scenarios demonstrates that lower integration levels (below 45\%, including the current level) exhibit minimal impact on system resilience in this event. However, surpassing this critical level without additional flexibility resources can exacerbate the failure probability due to substantially enlarged energy imbalances.
\end{abstract}

\vspace{1cm} 

Climate extremes, especially tropical cyclones — commonly known as hurricanes or typhoons — have threatened energy infrastructure over decades, leading to numerous widespread catastrophic blackouts globally\cite{xu2024resilience,stankovski2023power}. Despite the growing emphasis on enhancing the resilience of power systems, which are fundamental to socio-economic functioning, weather-associated power outages in the U.S. have escalated by 78\% during this decade compared to the last decade \cite{ClimateCentral2022}. Hurricane Maria (Category 5) in 2017 and Hurricane Fiona (Category 1) in 2022 plunged the entire island of Puerto Rico and its 1.5 million electricity customers into darkness\cite{USDeptEnergy2017, USDeptEnergy2022}, resulting in an estimated cost of US \$113.3 billion\cite{NOAAUSDollars,massol2018renewable}.
.

Meanwhile, in the context of long-term decarbonization roadmaps, various ambitious renewable energy integration targets have been set for power grids\cite{davis2018net,meadowcroft2023governing}. For instance, Puerto Rico has committed to achieving a 100\% renewable power grid by 2050\cite{PuertoRico2019}. The large-scale integration of variable renewable energy such as solar photovoltaic (PV) and wind energy notably increases the uncertainty in power system operations and decreases the grid inertia\cite{denholm2021challenges}. Especially, the integration of behind-the-meter (BTM) solar PV systems associated with unregulated individual-optimized storage, can further challenge grid operations due to unexpected demand fluctuations\cite{smith2022effect}. Moreover, these environment-dependent renewable sources are particularly sensitive and vulnerable to extreme weather events. Solar panels have been observed to exhibit greater fragility in storms than their design requirements\cite{ceferino2023bayesian}. Additionally, hurricanes associated with large cumulonimbi substantially reduce solar generation even prior to their landfalls\cite{ceferino2022stochastic}, which can further enlarge imbalances between electricity supply and demand. These challenges highlight the potential amplified risk of catastrophic blackouts due to climate-energy interactions and underscore the importance of developing a coupled climate-energy model to quantify the effects of climate extremes on the resilience of renewable power systems\cite{xu2024resilience,craig2022overcoming}.

Catastrophic blackouts triggered by initial common-cause disturbances have been extensively investigated. For instance, physics analyses reveal that small initial common-cause failures of critical lines can propagate through the transmission network, triggering catastrophic network collapse\cite{yang2017small,dobson2007complex,buldyrev2010catastrophic,gao2016universal}. A network cascade model demonstrates more severe failures when incorporating dynamics of power systems in contrast to the findings from static power flow analysis\cite{schafer2018dynamically}. Significant progress in understanding energy resilience has been widely achieved through efforts to bridge climate extremes and energy systems. Long-term coupled climate-energy planning for Sweden\cite{perera2020quantifying}, various EU cities\cite{perera2023challenges} and Puerto Rico\cite{bennett2021extending}, reveals the potential for catastrophic socio-economic consequences airing from extreme weather conditions. Towards the operation resilience of bulk power grid under specific events, the effectiveness of critical line hardening considering network cascades has been validated by synthetic Texas transmission grid under hurricanes\cite{sturmer2024increasing}. The hydrological-power cascade effect has been demonstrated in the California transmission grid during extreme events characterized by high-temperature stress\cite{webster2022integrated}.

Despite the substantial advances made in understanding the common-cause network cascade mechanism\cite{yang2017small,dobson2007complex,buldyrev2010catastrophic,gao2016universal,schafer2018dynamically} and the operation resilience\cite{sturmer2024increasing,webster2022integrated} of transmission grids under extreme events, two critical obstacles remain in quantifying spatiotemporal power outages during extreme weather events such as hurricanes (i) the lack of high-resolution spatiotemporal power outage models covering the comprehensive effect of evolving climate weather hazards on grid operations, transmission and distribution networks, and renewable generation; and (ii) a limited understanding of the effects of environment-dependent renewable penetration on cascading power outages. In practical blackout events under climate extremes, it is not only the failures of transmission networks but also those of more vulnerable and extensive distribution networks\cite{feng2022tropical}, as well as affected environment-sensitive renewable generation capacity\cite{ceferino2022stochastic}, that collectively contribute to these outages. However, due to the limited availability of comprehensive energy system data and high-resolution outage information of real-world events, existing research efforts on catastrophic blackouts have predominantly focused on cascading failures within the topology of transmission networks.

To address these challenges, we develop a coupled climate-energy model to capture and quantify the cascading power outages in renewable power systems under climate extremes. This model accounts for the comprehensive effects of evolving extreme weather conditions on renewable generation, and transmission and distribution networks. Our model is validated by a retrospective analysis of the 2022 Puerto Rico catastrophic blackout during Hurricane Fiona, a milestone event documented with the first-ever high-resolution spatiotemporal outage data on a weather-induced system-wide blackout. Using thousands of realizations that account for the uncertainty in grid infrastructure resilience, we investigate the resilience and vulnerability patterns of the grid under an evolving hazard. To further explore the role of distributed solar-dominated renewable integration in this catastrophic event, we then conduct a sensitivity analysis across a wide range of renewable integration levels. Beyond this specific event, our methodology offers a broadly useful tool for assessing the risks associated with different generation portfolios of a regional power system in response to projected climate extremes.

\begin{figure*}[b!]
\centering
\includegraphics[width=16.4cm]{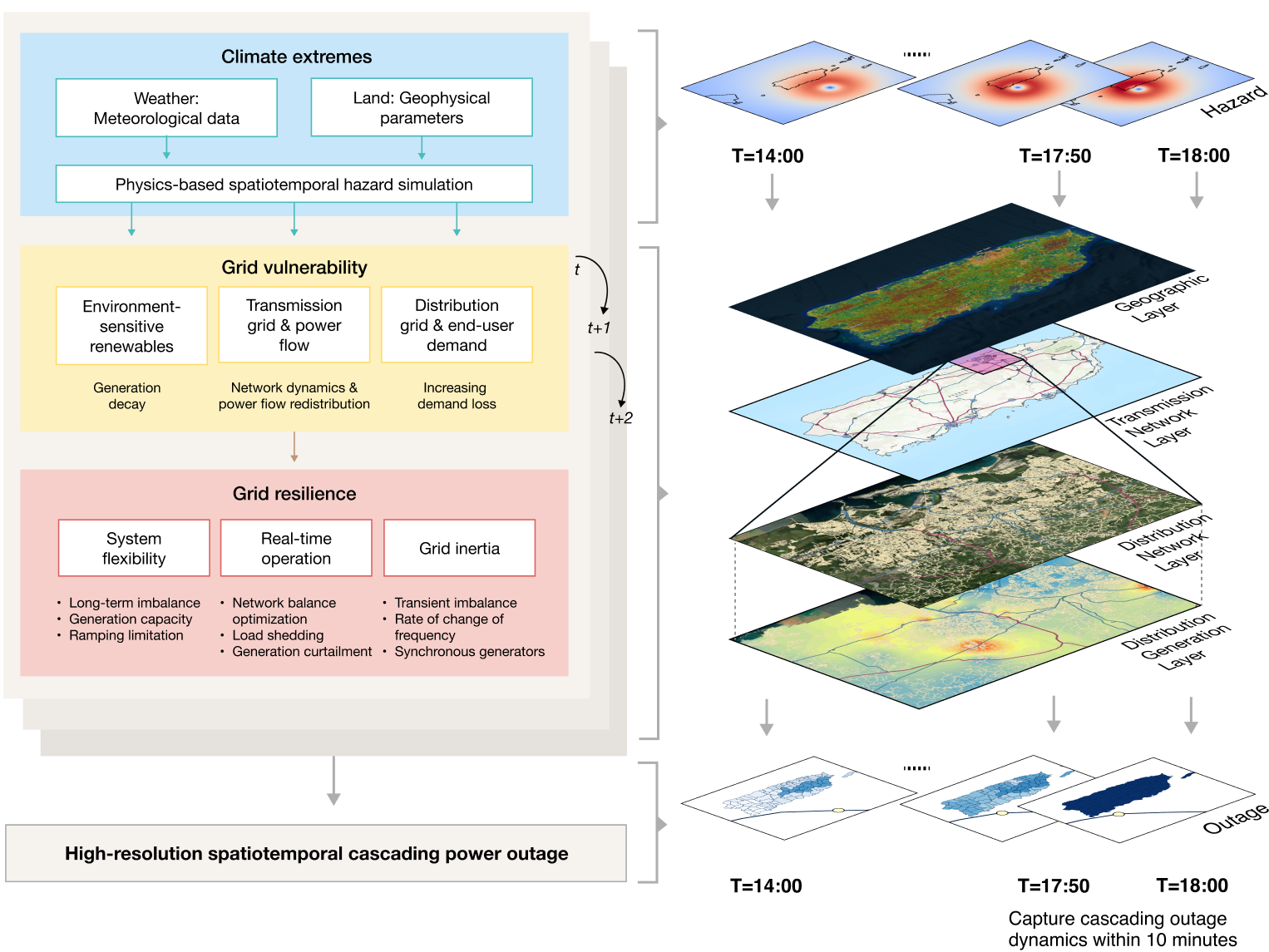}
\caption{\textbf{Schematic diagram of the CRESCENT model}. The proposed CRESCENT model enables the high-resolution spatiotemporal analysis of the climate extreme effect on comprehensive energy systems and captures the cascading outage dynamics.}
\label{fig1}
\end{figure*}

\section*{Results}

\subsection*{Modeling climate-induced power outages with renewable integration}
We propose a Climate-induced Renewable Energy System Cascading Event (CRESCENT) model. This model bridges the dynamics of renewable energy systems with evolving extreme weather events by integrating a spatiotemporal hazard model, a renewable energy system vulnerability model, and a multi-scale spatiotemporal cascading power outage model (see Fig. 1 and Methods). The meteorological component exemplifies tropical cyclones as a typical climate extreme to the Puerto Rico energy system and employs a physics-based tropical cyclone wind field model to generate high-resolution spatiotemporal wind hazards. The renewable energy system vulnerability model is used to simulate the effects of climate extremes on spatiotemporal infrastructure damage and renewable generation reduction. We assess the spatiotemporal infrastructure damage to both transmission and distribution networks through hazard resistance risk analysis (see Methods). The damage to distribution feeders within distribution networks results in the spatiotemporal loss of demand at the transmission network level, affecting grid operations. We consider the shutdown of wind turbines under extreme wind conditions and evaluate the climatic sensitivity of solar PV generation. This evaluation includes the analysis of generation decline in both utility-scale and distributed rooftop solar PV systems, attributed to wind-induced panel damage and reduced solar irradiance due to extensive cumulonimbus clouds. 

\begin{figure*}[b!]
\centering
\includegraphics[width=16.4cm]{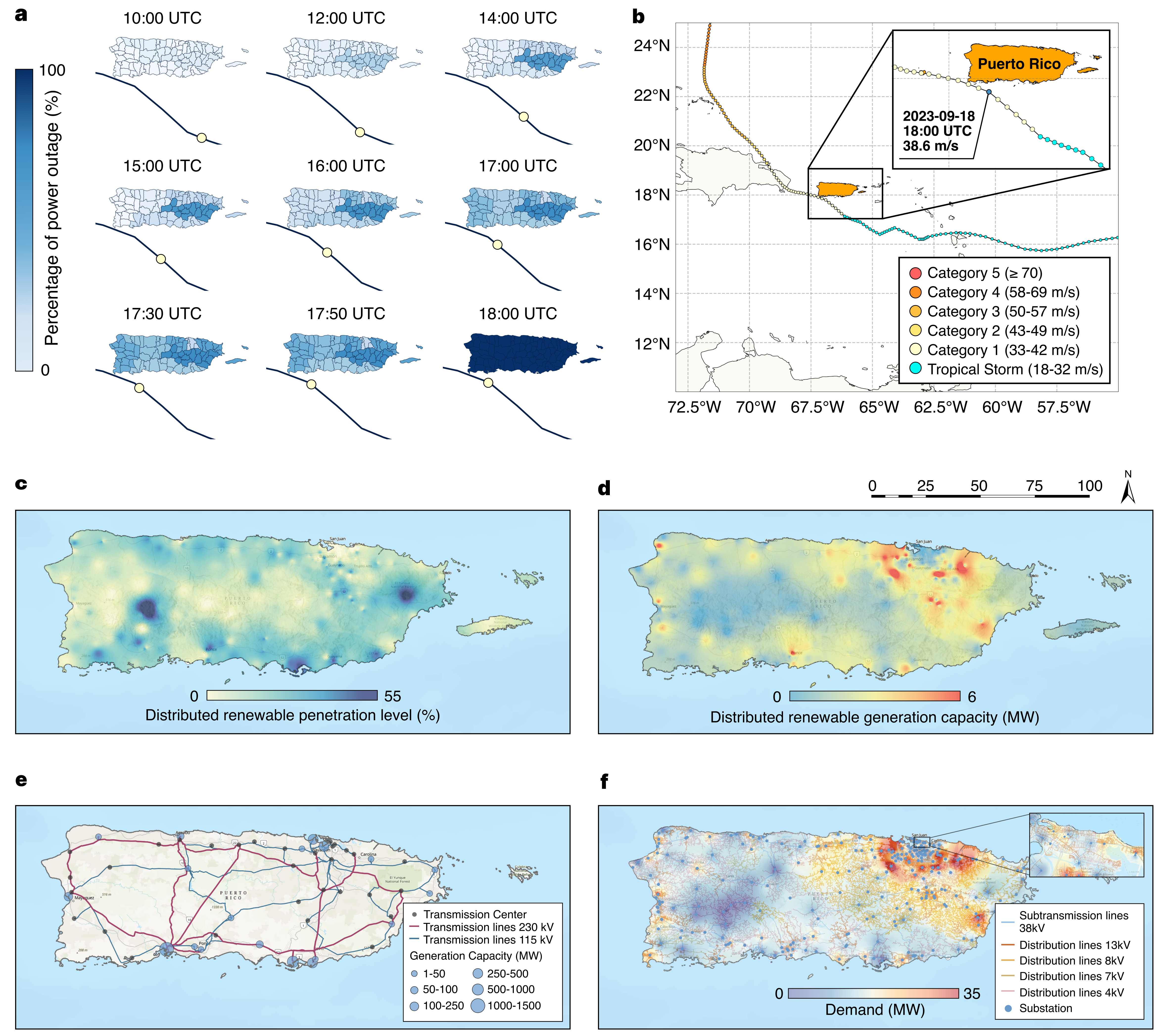}
\caption{\textbf{The 2022 catastrophic blackout in Puerto Rico during Hurricane Fiona and contemporary Puerto Rico renewable power system.} \textbf{a}, Spatiotemporal power outage map of the Puerto Rico power grid from 10:00 UTC to 18:00 UTC on September 18, 2022, along with the track (black line) and center location (yellow dot) of Hurricane Fiona. \textbf{b}, The track of Hurricane Fiona, highlighted by its maximum sustained wind speeds. Hurricane Fiona was a Category 1 hurricane during its landfall in Puerto Rico. \textbf{c-d}, Maps of (c) the penetration level and (d) the capacity of distributed renewable generation (rooftop solar systems; with data recorded at the distribution feeder level). \textbf{e}, Transmission network with generation capacity. \textbf{f}, Distribution network with the spatial distribution of demand. The power grid exhibits a supply-demand structure with major generation in the south and the primary load centers in the north (San Juan). The structure heightens the risk of power imbalance during disruptions of the transmission network. Despite certain distribution feeders reaching a renewable penetration level of 55\%, the overall penetration level with a total installed distributed renewable generation capacity of 296 MW remains below 20\% of the peak demand (2751 MW), approximately at 16\%.}
\label{fig2}
\end{figure*}

Informed by the vulnerability of renewable energy systems during evolving climate extremes, we propose a multi-scale spatiotemporal cascade model to quantify cascading power outages, featuring a temporal resolution compatible with the real-time operations of the decision-making system. Expanding the existing network cascade model that considers the overloaded line tripping and power flow redistribution induced by common-caused initial failures\cite{yang2017small,dobson2007complex,buldyrev2010catastrophic,gao2016universal,schafer2018dynamically}, the proposed method further accounts for the interaction between system resilience and evolving hazards. Under extreme weather conditions, system resilience/reliability is fundamentally determined by two key factors influencing power balance across different time scales: grid inertia for alleviating transient imbalances and system flexibility for eliminating sustained imbalances\cite{xu2024resilience}. Our cascade model considers power imbalances associated with renewable energy integration by taking these two factors into account. In the context of large-scale renewable integration, reduced grid inertia against transient power imbalances is considered in the network stability analysis. System flexibility considering reduced renewable generation and availability of flexible resources during climate extremes is embedded in real-time operation optimization (see Methods).

\begin{figure*}[!b]
\centering
\includegraphics[width=16.4cm]{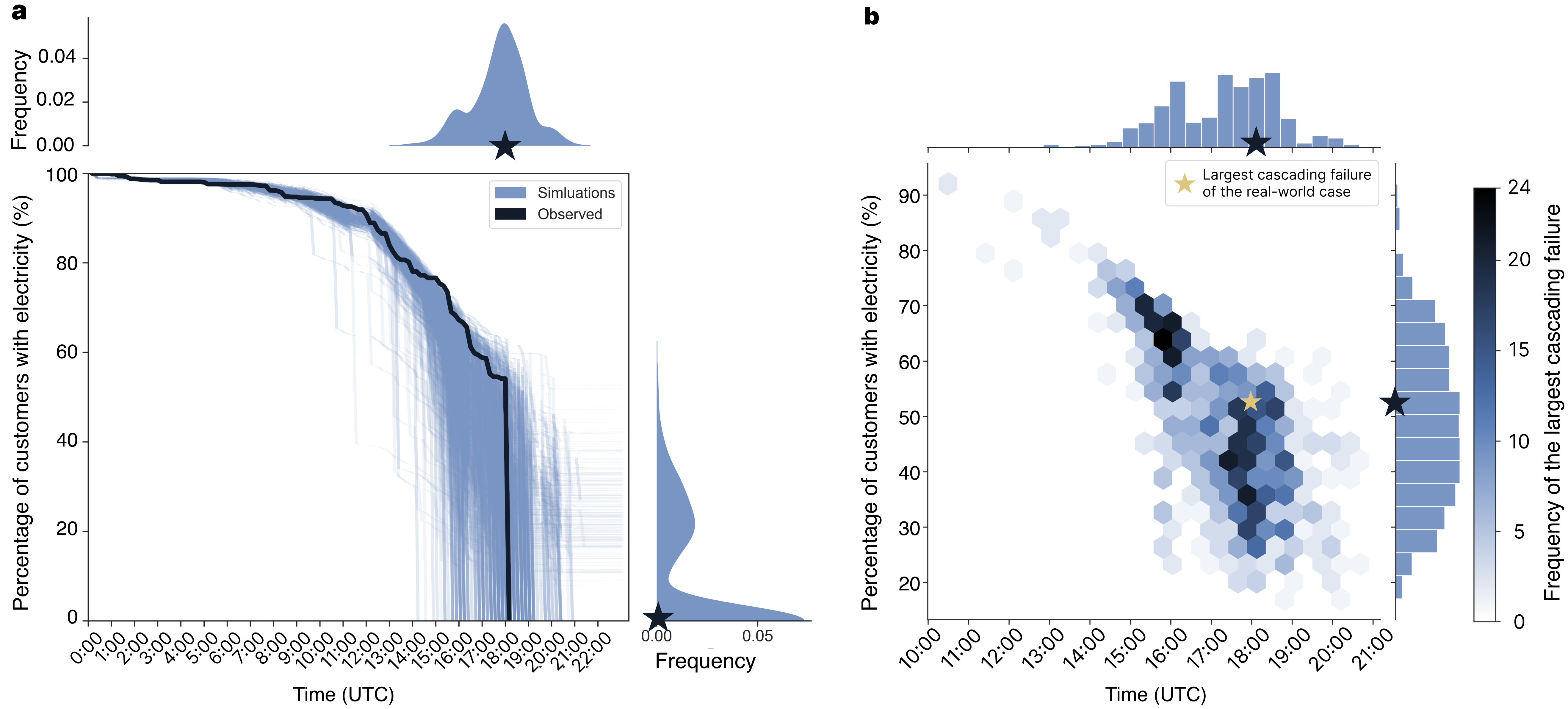}
\caption{\textbf{Realizations of cascading power outages}. 
\textbf{a}, Comparison of the percentage of total customers with electricity between the simulated and observed cases for the Puerto Rico power grid during Hurricane Fiona on September 18th, 2022. The 1000 simulated cases (blue) are generated from the proposed CRESCENT model with the contemporary grid configuration. The observed peak outage representing the degradation of customers with electricity (black) is obtained from the US power outage datasets\cite{Poweroutageus} recorded by local power utilities. The right-side subfigure shows the distribution of final system statuses (percentage of customers with electricity) across all simulations. The top subfigure shows the distribution of the times when catastrophic blackouts (100\% failure) occurred across all catastrophic blackout cases. \textbf{b}, Distribution of the largest failures. The largest failure in each power outage realization is identified by the largest drop in the percentage of customers with electricity between two successive simulation time steps. A darker color in a hexagon indicates a higher density (frequency) of data points within that area. The top and right histograms show the distributions for times and system statuses, respectively, where the largest cascading failures occurred. Black stars in the top and right distributions mark the position of the observed case.
}
\label{fig3}
\end{figure*}

\subsection*{Catastrophic blackout of Puerto Rico in 2022}
On September 18th, 2022, Hurricane Fiona plunged the island of Puerto Rico into complete darkness again after a similar situation caused by Hurricane Maria in 2017. During this system-wide blackout event, LUMA Energy, the local system operator that took over the grid in 2021, captured high-resolution spatiotemporal outage data\cite{Poweroutageus} in 10-minute intervals (i.e., at the real-time system dispatch scale). This marked the first recorded instance of such detailed data for a system-wide blackout caused by extreme weather events. Compared to regional-level power outage events, system-wide blackouts provide unique opportunities for exploring the weather-induced cascading failures and stability mechanisms of complex dynamic systems. This time-series outage data reveals a catastrophic cascading failure in the Puerto Rico power grid, with system outages escalating from below 50\% to 100\% within 10 minutes at 18:00 UTC, prior to the landfall of a mere Category 1 hurricane (Figs. 2a-2b). Consistently, the US official report\cite{USDeptEnergy2022}  indicates that escalating damages to distribution and transmission infrastructure led to a system-wide imbalance between electricity supply and demand. This imbalance triggered the off-grid protection of generation units, resulting in system instability.

We used the CRESCENT model to perform spatiotemporal simulations of the Puerto Rico power grid during Hurricane Fiona, with the compatible 10-minute temporal resolution as the power outage data and real-time operation. The energy-related models are configured based on the contemporary Puerto Rico power grid in 2022 with its utility-scale generation, distributed renewable generation, and transmission and distribution networks (Figs. 2c-2f), corresponding to the conditions during Hurricane Fiona that caused the latest catastrophic blackout. Considering the uncertainty in infrastructure resistance, we generated 1000 time-series power outage realizations, spanning from 0:00 to 23:00 UTC on September 18, 2022, in order to obtain meaningful simulation results. As shown in Fig. 3a, in the early stage (before 10:00 am UTC), characterized by a lower intensity of the hazard and higher functional integrity of the system, failures predominately occur within the distribution network, leading to a gradual degradation of system performance (characterized by the percentage of customers with electricity). However, as the hurricane approaches with increased hazard intensity, the distribution network sustains more severe damage and solar-based renewable generation reduces significantly. For instance, by 16:00 UTC (12:00 local time), distributed solar generation across the island declines to only 26.8\% of generation under clear sky conditions, with a reduction of over 125 MW generation (See Fig. S14). These factors can result in substantial supply-demand imbalances within a short timeframe (e.g., a 10-minute control cycle of system real-time operations). Under these conditions, system operation under limited flexibility may be unable to completely eliminate power imbalances, thus necessitating proactive load shedding and leading to a precipitous decline in system performance. Moreover, the grid faces a higher failure risk of transmission towers and lines under the intensifying hazard. Transmission line tripping, which transfers power flow to remaining lines, can lead to line overloads and potentially trigger cascading failures. In the worst-case scenario, insufficient grid inertia against substantial transient imbalances can destabilize the grid, resulting in a catastrophic blackout. Among all the realizations, 60\% result in catastrophic blackouts with complete system-wide outages. The occurrence of these catastrophic blackouts peaks at 18:00 UTC, the same time as for the real-world case.

Additionally, we identified the largest failure (the greatest decline in system performance) of each realization (Fig. 3b). In the majority of the realizations, the largest failures occur between 17:00 UTC and 19:00 UTC under 40~55\% system performance. During this period, the hurricane was closest to the island and generated the highest winds (Figs. S4-S10). The system had been progressively weakened by cumulative damages, including less robust topology connection and decreased system flexibility and grid inertia. Subsequent failures could then directly devastate the power grid. These findings closely align with the real-world catastrophic blackout incident, highlighting a high-risk state for a weakened system. The second peak of the largest failures is at a relatively early stage, between 15:00 UTC and 16:00 UTC. The early timing of the largest failures in the system indicates that critical components exist, and these failures correspond to scenarios in which critical components were assigned lower resistance on the hazard resistance distribution of components. Failures of critical components could contribute to severe degradation of the system, even at the early stage when the hazard is less severe and the system is not yet significantly weakened.

\begin{figure*}[!b]
\centering
\includegraphics[width=16.4cm]{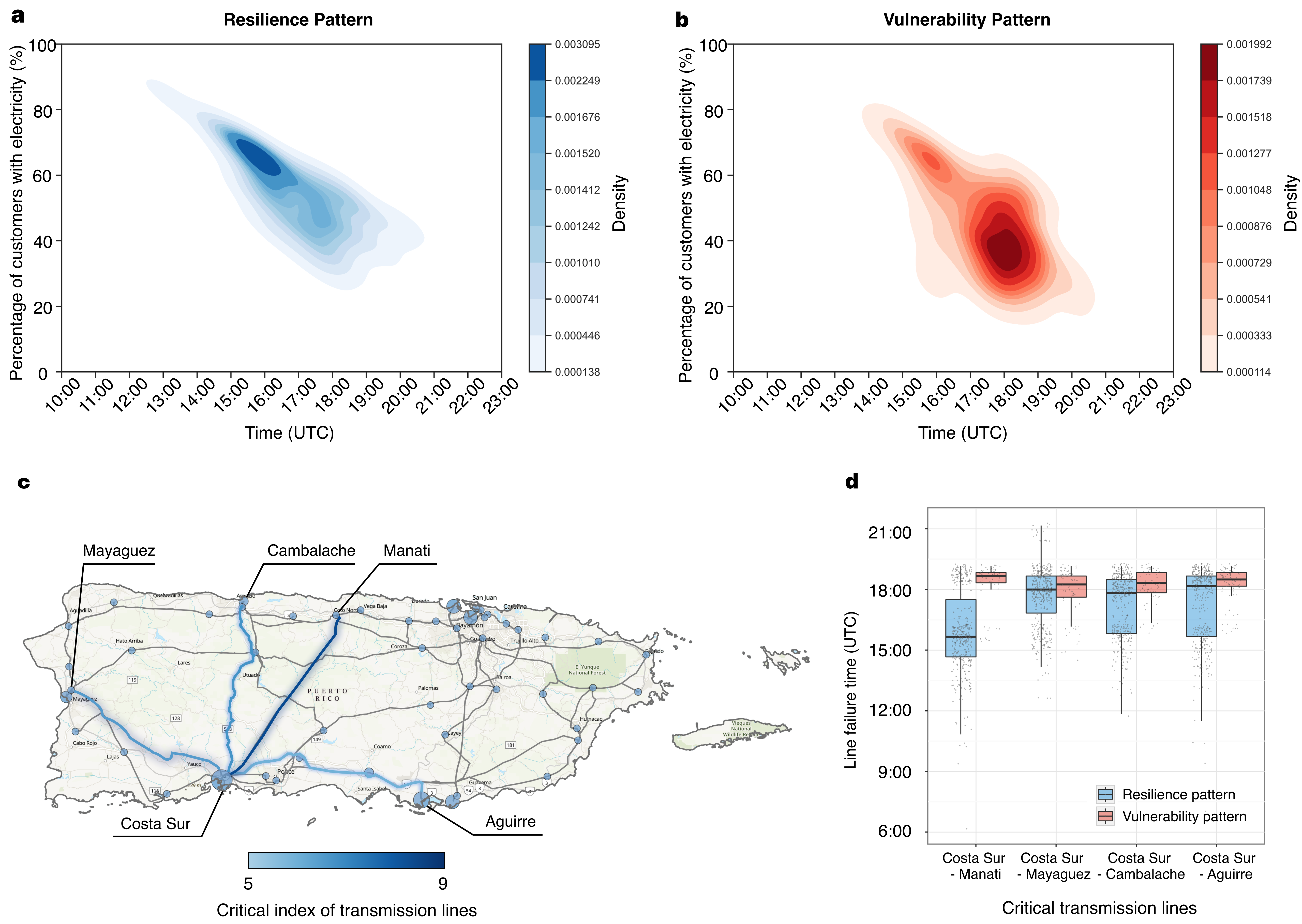}
\caption{\textbf{Resilience and vulnerability patterns}. \textbf{a}, Resilience pattern. The kernel density estimation of the largest failure (the most significant drop in system performance) in each resilient case (where the grid survives without a complete blackout) among the 1000 realizations. The darker color (blue) represents areas where the largest failures occur in the majority of resilient cases. \textbf{b}, Vulnerability pattern. Similar to (a) but for vulnerable cases (where the grid suffers a catastrophic blackout) among the 1000 realizations. \textbf{c}, Identified critical transmission lines in the transmission network. The critical index for a transmission line is defined as the proportion of instances where its failures directly contribute to catastrophic blackouts (a 100\% complete outage occurs as soon as the line fails) across all realizations. Only those lines with a critical index higher than 5\% are highlighted (blue), while the remaining lines are colored in grey. The names of substations associated with these critical lines are labeled. \textbf{d}, Resilience and vulnerability patterns of the top four critical transmission lines. The resilience pattern boxplots (blue) show the distributions of line failure times in instances where the system ultimately remains functional despite the failure of the corresponding lines. The vulnerability pattern boxplots (red) show similar distributions in instances where a catastrophic blackout occurs immediately following the failure of the specific lines. Tiny grey dots represent individual data points. The horizontal black line within each box indicates the median, box edges show the interquartile (50\%) range, and whiskers extend to the 5th and 95th percentiles.
}
\label{fig4}
\end{figure*}

\subsection*{Resilience patterns}

A complete power outage (catastrophic blackout) is far more severe and demanding than an extensive power outage with a functional grid, as it requires substantial effort to restore the system-level balance and synchronization. To further explore system resilience against this extreme event, therefore, we categorized all the realizations into resilient (without complete power outages) and vulnerable (with complete power outages) sets (Figs. 4a-4b). It is noted that the resilient cases experience their largest failures significantly earlier than the vulnerable cases. This result suggests that systems that experience early severe failures but maintain adequate functionality (e.g., above 50\% system performance) can be more robust to ride through subsequent damages under increasing hazards. This finding indicates that passive early degradation might enhance system resilience. This resilience pattern, previously unidentified, shares a similar philosophy to proactive de-energization measures such as Public Safety Power Shutoffs\cite{wang2023local} used in California against wildfires by reducing the grid’s burden and alleviating the magnitude of energy imbalances. 

The disparity in resilience and vulnerability patterns observed from a system-level perspective further motivates our considerations of how the failure timing of individual components, especially those with the most impact, affects system resilience. To identify these critical components, we define the critical index of a transmission line as the proportion of instances where its failures directly contribute to catastrophic blackouts among all realizations. The top four critical lines (Fig. 4c), all connected to Costa Sur — the largest power plant complex in Puerto Rico, play a vital role in the connectivity of transmission network topology, such as delivering electricity to the north region and establishing connections with the second-largest power plant complex (Aguirre). Regardless of this specific weather event, the steady-state topology analysis based on the current flow between centrality also suggests Costa Sur has the highest importance for the grid (see details in Supplementary Note 3).

The resilience patterns of these critical transmission lines (Fig. 4d) suggest that earlier failures of these critical lines (when the system maintains relatively adequate functionality) tend to enhance overall system resilience, consistent with the resilience patterns at the system level (Fig. 4a-4b). In a later stage during the extreme event, the system is compromised in network connectivity, and the disruptions of critical lines followed by the large-scale power flow redistribution can overload the remaining lines, which may result in further overloads and incur cascading failures. Moreover, in a weakly connected network, the removal of these critical lines that connect to the highest-centrality node could lead to network segmentation, causing substantial power imbalances in the sub-networks and potentially leading to instability. To further validate this finding, we preset the hazard resistance of the most critical component (Costar Sur – Manati) to be either the weakest or the strongest among all components. When this critical component ranks within the weakest 1\% and 10\%, the probability of catastrophic blackouts under this extreme event is reduced by 17\% and 9\%, respectively. While scenarios featuring the critical component with higher wind hazard resistance exhibit increased failure probability (See Fig. S13).

\subsection*{Sensitivity analysis of increasing renewable integration}

To explore the effects of increasing renewable integration, especially unregulated BTM distributed solar PV systems, on the risk of catastrophic blackouts, we conduct a sensitivity analysis using the proposed CRESCENT model under the same hurricane event. Here, we generated 1000 realizations considering the uncertainty in infrastructure resilience for each of the renewable integration levels ranging from 10\% to 80\%, where the renewable integration level is defined as the proportion of demand met by solar PV systems. Despite some BTM solar systems being equipped with energy storage for consumer self-sufficiency, such systems are not well regulated and dispatched by the grid operators and can even undermine grid resilience due to their individual optimal strategies\cite{smith2022effect}. Therefore, to simplify this sensitivity analysis, they are configured without additional storage and set to operate in the widely adopted maximum power point tracking (MPPT) mode\cite{subudhi2012comparative}. To eliminate regional variations, we proportionally adjusted the renewable integrations according to the demand profiles of distribution feeders, ensuring a consistent level across the entire island.

\begin{figure*}[!b]
\centering
\includegraphics[width=15cm]{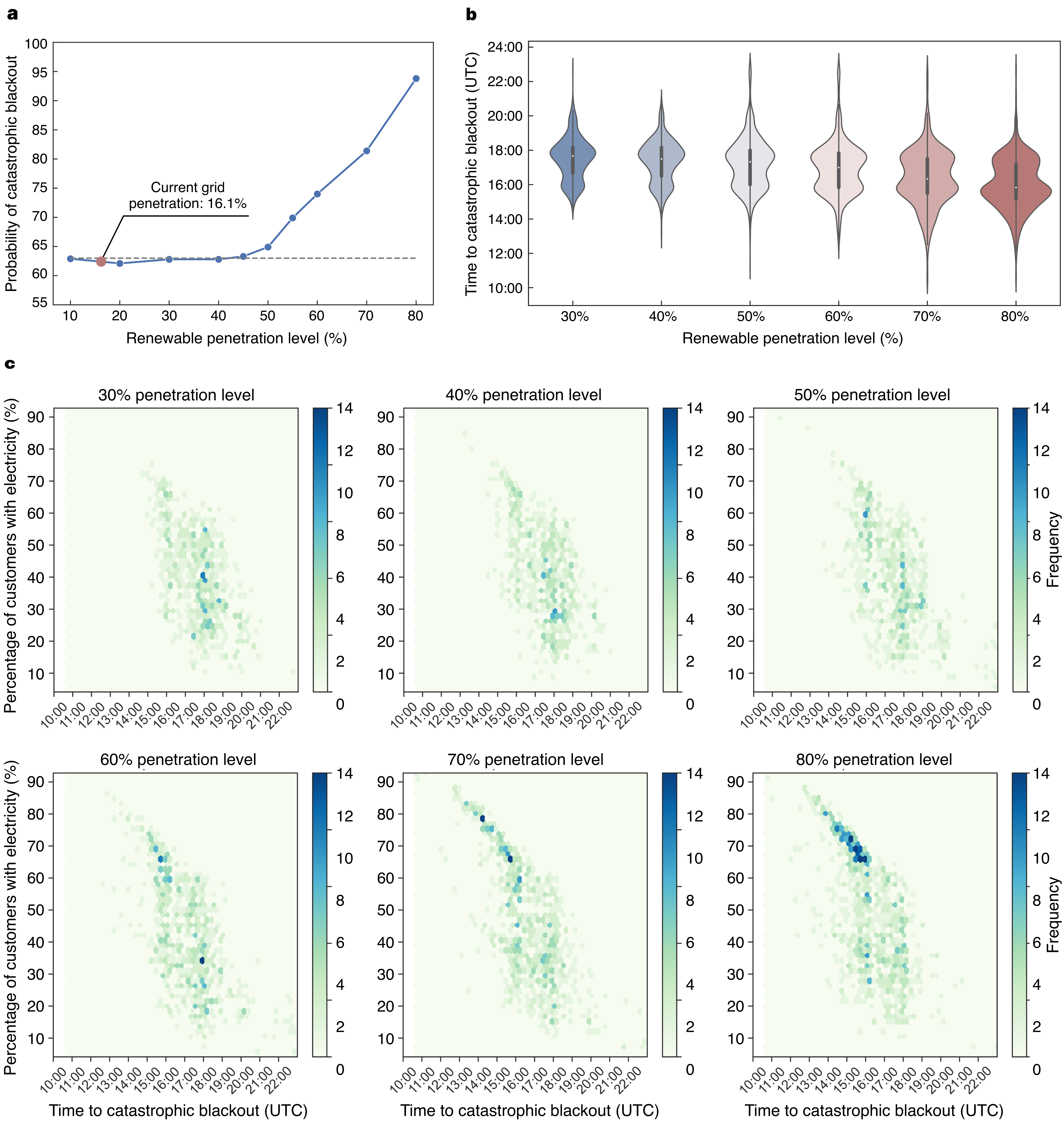}
\caption{\textbf{Sensitivity analysis of renewable energy integration on catastrophic blackouts under the same hurricane event}. \textbf{a}, Probability of catastrophic blackout occurrence at varying levels of renewable energy integration. The current grid (contemporary during Hurricane Fiona) with an average renewable integration level of 16.1\% is marked by the red dot. textbf{b}, Violin plots for time distributions of catastrophic blackouts at varying levels of renewable energy integration. textbf{c}, Distributions of catastrophic blackout occurrences at different renewable integration levels. Each data point in the plots records the time of a catastrophic blackout occurrence and the corresponding system status (percentage of customers with electricity). A darker color in a hexagon indicates a higher density (frequency) of data points within that area. The probability for each blue point in (a) and the violin plots in (b) were calculated based on 1000 realizations generated by the proposed CRESCENT model at a specific renewable integration level under Hurricane Fiona. In the violin plot, each violin represents the distribution of blackout occurrence times at a specific level of renewable energy integration. The width of each violin indicates the frequency (probability) of blackouts at different times. The black bar within each violin shows the interquartile range, and the inside white dot represents the median time of catastrophic blackout occurrence. A larger violin size in (b) indicates more catastrophic blackouts.}
\label{fig5}
\end{figure*}

By comparing these realizations, we observe a nonlinear effect of increasing environment-sensitive renewables on potential catastrophic blackouts (Fig. 5). Below a renewable integration level of approximately 45\%, including the current level of 16.1\%, realizations exhibited similar probabilities of catastrophic blackouts and nearly identical failure patterns. This phenomenon suggests that below this level, the risk introduced by the growth of solar generation has a minimal impact on system resilience. This can be attributed to the real-time energy balances between supply and demand still being led by adequate conventional generation, which is resilient to extreme weather impacts.

However, as the renewable integration level in the system further increases, the effects of renewables begin to emerge with the failure probability exhibiting a super-linear growth (Fig. 5a). Under a high renewable integration level, the extreme weather event significantly diminishes solar-dominated generation due to wind damage and cumulonimbus cloud cover, which enlarges energy imbalances between supply and demand and further challenges grid inertia and flexibility for maintaining system stability. Therefore, the system becomes even more likely to experience a catastrophic blackout in an earlier stage (Fig. 5b). 

The temporal distribution of catastrophic blackouts under various renewable integration levels (Fig. 5c) more clearly demonstrates that in a renewable-dominated scenario with over 70\% renewable integration, grids face significant challenges in managing even the lower intensity of hazards at early stages, despite having less cumulative damage and relatively higher system functionality (above 65\%). Conversely, under the same hazard, grids with lower renewable integration (below 50\%) exhibit higher resilience during the initial phases. This difference is primarily attributable to the substantial reduction in power generation of the solar-dominated power system caused by the indirect effect (cloud cover) of hurricanes on such environment-sensitive renewables (see Fig. S15), which exacerbates energy imbalances.

\section*{Discussion}

In the context of extreme weather events like hurricanes that cause extensive damage to electric power systems, our study presents the CRESCENT model to analyze cascading failures in climate-induced power outages. This model distinguishes itself from the existing cascading outage models at the transmission network level\cite{yang2017small,dobson2007complex,buldyrev2010catastrophic,gao2016universal,schafer2018dynamically,sturmer2024increasing} by accounting for the spatiotemporal climate-energy dynamics of renewable power systems. It comprehensively incorporates the effects of evolving climate extremes on utility-scale and distributed generations, and transmission and distribution networks. This model also stands out by using the real-world grid datasets, validated by a weather-induced catastrophic blackout with the first-ever high-resolution outage data records (2022 Puerto Rico blackout during Hurricane Fiona). The CRESCENT model is adaptable to various regions and can be further employed to analyze the resilience of energy transitions in future climates, by incorporating storms projected in climate change scenarios\cite{xi2023increasing}. Also, our study exemplifies hurricanes as a primary climate threat to Puerto Rico; in the future CRESCENT can be extended to model the impact of other climate extremes (e.g., flooding and heatwaves) to assess the overall climate resilience of renewable energy systems.

In analyzing thousands of model-generated realizations for the Puerto Rico power system during Hurricane Fiona, we identified distinct resilience and vulnerability patterns. These patterns reveal that early degradation of the system or early failure of critical components can counterintuitively enhance system resilience, helping the grid ride through subsequent hurricane-induced damages. This paradoxical effect can be explained by the mechanism that early failure can reduce the energy burden on the grid and even separate the grid into multiple sub-grids in advance. Such early degradations can also alleviate the transient power imbalances and prevent the propagation of cascading failures in a later phase. Although proactive grid regulation strategies such as proactive power shutoff\cite{wang2023local} and organized microgrid operation\cite{jin2021building} are recognized to be effective, our analysis reveals that even passive failures occurring early can also enhance grid resilience against evolving climate extremes. Moreover, our model can be designed to support these proactive climate-resilient strategies, offering assistance in their implementation.

To explore the role of renewable integration in this catastrophic blackout, we perform a sensitivity analysis across various solar-dominated distributed integration levels ranging from 10\% to 80\% within the context of the same event. Our results show the nonlinear impact of renewable energy integration on system resilience for Puerto Rico during the same event. Below an approximately 45\% renewable integration level, including the current level of 16.1\%, the integration of distributed solar systems is well-accommodated without compromising system stability. In contrast, surpassing this level towards a 100\% renewable grid may significantly heighten the risk of climate-induced cascading power outages, primarily due to enlarged energy imbalances resulting from substantial reductions in renewable generation that further challenges the grid inertia and system flexibility.

We note that this sensitivity analysis does not account for energy storage associated with behind-the-meter solar installations due to the current lack of a unified dispatch mechanism for such individually optimized distributed systems, which are even recognized to potentially compromise grid resilience\cite{smith2022effect}. However, the coordinated aggregation of energy storage systems, providing virtual inertia and additional flexibility, presents a promising solution to mitigate the risks associated with the large-scale integration of renewable energy. While the quantitative results of the sensitivity analysis are particularly derived for this event, our methodology can serve as a broadly useful tool for assessing the risks associated with different generation portfolios of a regional power system in response to forecasted or projected extreme weather events. This enables stakeholders and grid operators to make informed decisions on optimizing generation portfolios and operation strategies for enhancing climate resilience, thereby ensuring a more sustainable energy transition. In future work, this methodology can be extended to investigate the potential heterogeneity in the impact of renewable energy integration on catastrophic blackouts across projected extreme events in a changing climate.

\section*{Methods}
\subsection*{Energy system configurations}
The configuration of the realistic Puerto Rico electric power system in 2022 (the period contemporary with Hurricane Fiona) is comprehensively defined by high-resolution datasets across four parts: utility-scale and distributed generation resources, transmission and distribution networks, and demand profiles.

Data on the capacity and detailed location information of all utility-scale generation units, including utility-scale solar PV systems, wind turbines, hydropower, and other traditional power plants as of September 2022 were obtained from U.S. Energy Information Administration (EIA) Form EIA-860 Preliminary Monthly Electric Generator Inventory \cite{USEnergyInfoAdmin2023}. Operation availability for all Puerto Rico power plants during Hurricane Fiona was obtained from the Daily Generation Availability Report (September 18th, 2022, the same day as Hurricane Fiona's landfall) of the local power utility (LUMA Energy, also the grid operator) \cite{LUMAEnergy2023}. Generation unit parameters (e.g., ramp up/down rates in MW/minute) were obtained from the Puerto Rico Integrated Resource Plan 2018-2019 for Puerto Rico Electric Power Authority (PREPA) \cite{PRIRP}. The installed capacity of distributed generation, specifically rooftop solar PV systems, of each distribution feeder in September 2022 was collected from the local power utility \cite{LUMAEnergy2023}. Geographic information system data for the Puerto Rico transmission and distribution networks, as recorded in September 2022, were also obtained from the local power utility \cite{LUMAEnergy2023}  (see visualizations in Figs. 2c-2f). This dataset includes segment types and parameters of all 115 kV and 230 kV transmission lines of the transmission network, which are subsequently used in calculating network cascades and steady-state power flow. The distribution network part of this dataset includes the detailed geospatial topology of 973 operational distribution feeders with their connections to substations (see Fig. 2f and Fig. S2). The demand profile of each distribution feeder is also included in this dataset. Other detailed configurations of the Puerto Rico power system, including generator parameters and illustrations of the transmission network, the distribution networks, and the distributed feeders, are shown in Table S1 and Figs. S1-S2.

\subsection*{Hurricane hazard model}
The track of Hurricane Fiona was sourced from the International Best Track Archive for Climate Stewardship (IBTrACS) at the National Center for Environmental Information of the National Oceanic and Atmospheric Administration (NOAA) \cite{schreck2014impact}, which includes time-series data of hurricane’s center locations, maximum sustained wind speed, and radius of maximum wind. To align with the temporal scale of real-time power system operations, we interpolated the track data from a 3-hour to a 10-minute interval. A physics-based wind profile model that accounts for both the tropical cyclone inner core and outer radii dynamics \cite{chavas2015model} is used to generate the hurricane wind profile according to the track information. By integrating this boundary-layer hurricane wind profile with environmental background flow derived from the storm’s translation speed \cite{lin2012hurricane}, the asymmetric spatiotemporal wind fields for the hurricane are obtained. The accuracy of this wind field model has been validated for simulating tropical cyclone hazards such as coastal winds, rainfall, and storm surges \cite{xi2020evaluation,wang2022investigation}. To further consider the land roughness impact on the surface-layer wind speed, we convert the wind speeds based on the land cover classes using the logarithmic vertical wind speed profile \cite{stull2012introduction,ro2007characteristic}. Data of Puerto Rico’s land cover classes, e.g., residential, forests, crops, and wetlands, were obtained from the National Land Cover Database (NLCD) of the United States Geological Survey (USGS) \cite{homer2020conterminous}. For detailed information on the land roughness of Puerto Rico, see Fig. S11.

\subsection*{Power outage data}
Spatiotemporal power outage data on the 2022 Puerto Rico blackout during Hurricane Fiona (as shown in Fig. 2a and the observed curve in Fig. 3a) were sourced from the US power outage datasets \cite{Poweroutageus}. The data for this specific event, detailing the percentage of customers without electricity in seven regions ( Fig. S3) of Puerto Rico, were recorded by the Puerto Rico grid operator, LUMA Energy \cite{LUMAEnergy2023}, at 10-minute intervals, spanning from pre-event to post-event periods. The island’s power grid experienced a system-wide catastrophic blackout between 17:50 and 18:00 UTC, September 18th, 2022. The data recording was terminated after the catastrophic blackout and was not recovered until September 20th, 2022. These outage data represent the first high-resolution spatiotemporal record of a system-wide blackout induced by a climate extreme event, as the grid’s modernization and digitalization have been improved by LUMA Energy since it became the grid operator in 2021.

\subsection*{Renewable power system vulnerability model}
The vulnerability model is used to generate spatiotemporal disturbances in the renewable power system for each realization. It accounts for infrastructure damage caused by hurricane winds and the decline in generation from environment-sensitive renewable energy resources. Infrastructure damage depends on the fragility of grid components, which include renewable energy structures (primarily solar panels), transmission towers, transmission lines, and distribution feeders. The damage estimates of these components are based on a series of functions known as fragility curves \cite{ceferino2023bayesian,bennett2021extending,dos2022methodology,xu2024hazard}, which describe the relationship between component failure probability and wind intensity. The selected fragility curves were particularly designed and calibrated for Puerto Rico, including utility-scale and distributed rooftop solar panels \cite{ceferino2023bayesian}, transmission lines \cite{bennett2021extending}, transmission towers \cite{dos2022methodology}, and distribution feeders \cite{xu2024hazard}. Applying existing fragility curves, which are mostly independent of hazard duration, directly in spatiotemporal risk analysis can lead to an overestimation of infrastructure damage due to repeated sampling \cite{xu2024hazard}. Therefore, we define resistance functions by taking the inverse of these fragility functions, thereby characterizing the distribution of wind resistance for each type of component. Rather than sampling fragility curves for time-varying failure probabilities of components, we assign time-invariant hazard resistances to components by sampling their resistance distributions in each realization. During the spatiotemporal risk analysis, if the wind intensity at a grid component exceeds its assigned resistance value, which is sampled at the onset of each realization, the component is considered to have failed. The spatiotemporal vulnerability model for generating disturbances is configured with a 10-minute time interval, aligning with the time scale of real-time operations.

The decline in renewable energy resources accounts for the spatiotemporal impact of hurricane-induced cumulonimbus cloud cover on solar irradiance, leading to reduced generation output from both utility-scale and distributed rooftop solar PV systems. The solar generation reduction rate is derived from a solar irradiance decay model \cite{ceferino2022stochastic}, validated by large-scale historical global horizontal irradiance data and Atlantic hurricane activity from the Atlantic hurricane database from 2001 to 2017 (see details in Supplementary Note 7). Also, as the wind intensity (above 33 m/s) of a hurricane exceeds the typical cut-off speed (25 m/s) of existing wind turbines, wind turbines are considered to shut down during the hurricane period.

\subsection*{Multi-Scale Spatiotemporal Cascade Model}
Informed by the renewable power system vulnerability model under evolving climate extremes, a multi-scale spatiotemporal cascade model is developed to simulate network dynamics. We expand state-of-the-art network cascade models designed for common-caused initial disturbances, to account for the decision-making system’s network dynamic evolution influenced by multi-scale system resilience factors — grid inertia and system flexibility — under climate extremes.

The system’s dynamic evolution is discretized based on the real-time operations of the power system in a 10-minute time interval with a temporal discretization set denoted by $\mathcal{T}$. At each time step $t \in\mathcal{T}$, given the damage state estimated from the hazard and vulnerability models, we conduct a topology analysis of the transmission network to identify a set $\boldsymbol{\mathcal{G}}\left(t\right)$ for all connected subgraphs $\mathcal{G}_n\left(t\right)$, that is, functional sub-grids with active generators and demand nodes. For each functional sub-grid, the network cascading dynamics are built upon the widely adopted ORNL-PSERC-Alaska (OPA) cascade model\cite{dobson2007complex,bialek2016benchmarking}. This model simulates the process of line overloads and subsequent cascading line tripping, triggered by initial failures due to infrastructure damage and subsequent power flow redistribution. The network cascading failures could reform the network topology with an updated set $\boldsymbol{\mathcal{G}}^\prime\left(t\right)$. Within a functional sub-grid $\mathcal{G}_n^\prime\left(t\right) \in \boldsymbol{\mathcal{G}}^\prime\left(t\right)$, the power imbalance $\Delta P_{imb}^{\left(n\right)}\left(t\right)$ of this sub-grid is determined by the deviation of the total generation $\sum_{g\in\mathcal{P}_n} P_{g,t}$ (accounting for the decline in renewable energy resources during the extreme event) from total demand $ \sum_{d\in\mathcal{D}_n} L_{d,t}$ within the network, i.e.,

\begin{align}
  \Delta P_{\mathrm{imb}}^{\left(n\right)}\left(t\right)=\sum_{g\in\mathcal{P}_n} P_{g,t}-\sum_{d\in\mathcal{D}_n} L_{d,t},t\in\mathcal{T} \numberthis \label{eq1} 
\end{align}

\noindent where $\mathcal{D}_n$ and $\mathcal{P}_n$ denote the sets of demand and generation units within the sub-grid.

Grid inertia quantifies the system’s capability to mitigate the effect of the power imbalance. We use the rate of change of frequency (RoCoF) \cite{azizi2020local} jointly determined by the power imbalance and grid inertia provided by synchronous generators as a system stability constraint. A high RoCoF indicates that the system may reach a frequency nadir or zenith exceeding the system’s tolerance, thereby triggering the off-grid protection of generation units. We select the maximum RoCoF threshold as ±2 Hz/s following the recommendations by the National Renewable Energy Laboratory (NREL) for the Puerto Rico power grid \cite{gevorgian2019interconnection}. For each sub-grid at time step $t\in\mathcal{T}$, we calculate the maximum RoCoF (Hz/s) by

\begin{align*}
  \mathrm{RoCo}\mathrm{F}_{\mathrm{max}}^{\left(\mathrm{n}\right)}=\frac{f^0\Delta P_{\mathrm{imb}}^{\left(n\right)}\left(t\right)}{\sum_{g\in\mathcal{P}_n^{SG}}{2H_{SG,g}P_{SG,g}^{nom}}} \numberthis \label{eq2} 
\end{align*}

\noindent where $f^0$ is the rated 60 Hz frequency for the grid, $\mathcal{P}_n^{SG}$ denotes the set of synchronous generators, $H_{SG,g}$ and $P_{SG,g}^{Nom}$ represent the inertia constant and nominal generation capacity of a synchronous generator, respectively. The sub-grid is removed from the functional network set if its maximum RoCoF exceeds the ±2 Hz/s threshold. For a detailed derivation of the maximum RoCoF, see Supplementary Note 5. 

For surviving sub-grids, we further embed the decision-making process of the system’s real-time operations to eliminate power imbalances, accounting for system flexibility. The progressive decline in climate-sensitive renewable generation under evolving climate extremes requires more flexible resources, e.g., dispatchable generation units, to meet the power balance constraint of the network. The decision-making operations of the power system during climate extreme events are formulated by solving a mixed-integer linear programming (MILP)-based grid operation model, which is adapted from unit commitment (normal operation) and optimal load shedding (remedial control) models \cite{knueven2020mixed,aminifar2012impact}. The objective function of the grid operation model is designed to minimize the losses of load shedding and the curtailment of generation units by using flexible dispatchable resources within the network. The detailed model setup and programming solver for the grid operation model are provided in Supplementary Note 6.

\subsection*{Data and Code Availability}
The data and code will be made publicly available upon publication.

\subsection*{Acknowledgments}
L.X., N.L. and D.X. were supported by US National Science Foundation grant number 2103754 (as part of the Megalopolitan Coastal Transformation Hub) and Princeton University Metropolis Project. H.V.P. was supported in part by a grant from the C3.ai Digital Transformation Institute.

\subsection*{Author contributions}
L.X. contributed to conceptualization, methodology, writing of the initial draft. N.L contributed to the conceptualization, writing, review, editing, supervision, and guidance. H.V.P. contributed to writing, review, editing, supervision, and guidance. D.X. contributed to methodology and editing. A.T.D.P contributed to writing, review and editing.

\subsection*{Competing interests}
The authors declare no competing interests.

\clearpage 

\bibliographystyle{ieeetr}  
\bibliography{CascadingPaperArXiv}

\end{document}